\documentclass[useAMS,usenatbib]{mn2e}

\usepackage{amssymb}
\usepackage{amsmath}
\usepackage{times}
\usepackage{graphicx}

\newcommand{\lya}{Ly$\alpha$}

\newcommand{\msun}{\mbox{$M_\odot$}}

\newcommand{\kms}{\mbox{km s$^{-1}$}}
\newcommand{\cm}{cm$^{-2}$}
\newcommand{\civ}{\hbox{C\,{\sc iv}}}

\newcommand{\oi}{\hbox{O\,{\sc i}}}

\title[Intergalactic \civ\ absorption at redshifts 5.4 to 6]{Intergalactic 
\civ\ absorption at redshifts 5.4 to 6}

\author[Ryan-Weber, Pettini \& Madau]{Emma
 V. Ryan-Weber,$^{1}$\thanks{email: eryan@ast.cam.ac.uk} Max
 Pettini$^1$ and Piero Madau$^2$\\ $^1$Institute of
 Astronomy, Madingley Rd, Cambridge, CB3 0HA, UK\\ $^2$Department of
 Astronomy and Astrophysics, University of California, Santa Cruz, CA
 95064, USA}

\begin{document}

\date{Accepted 2006 June 28. Received 2006 June 28; in original form
2006 June 02}

\pagerange{\pageref{firstpage}--\pageref{lastpage}} \pubyear{2006}

\maketitle

\label{firstpage}

\begin{abstract}
We report the discovery of a strong \civ~$\lambda\lambda
1548,1550$ absorption system at $z_{\rm abs} = 5.7238$ in the
near-infrared spectrum ($J$-band) of the $z_{\rm em} = 6.28$ QSO SDSS
J1030+0524.  These observations, obtained with the Infrared
Spectrometer And Array Camera (ISAAC) on the European Southern
Observatory Very Large Telescope (ESO VLT), demonstrate that, with
modern instrumentation, QSO absorption line spectroscopy can be
successfully extended to near-infrared wavelengths to probe the
intergalactic medium near the end of the reionization epoch.  Although
the statistics of this pilot study are limited, the mass density of
triply ionized carbon implied by our data is comparable to the values
of $\Omega_{\rm C\,IV}$ reported at lower redshifts. Neither the
column density distribution of \civ\ absorbers nor its integral show
significant redshift evolution over a period of time which stretches
from 1 to 4.5\,Gyr after the big bang, suggesting that a large
fraction of intergalactic metals may already have been in place at
redshifts above 6. Alternatively, the strong \civ\ system we have
detected may be associated with outflowing, highly-ionized, gas from a
foreground massive galaxy; deep imaging and spectroscopy of galaxies
near the QSO sightline should be able to distinguish between these two
possibilities.
\end{abstract}


\section{Introduction}

A major goal of extragalactic astronomy is to identify the end of the
`dark ages' when the first luminous objects were assembled and the
Universe was reionized. It is an early generation of extremely
metal-poor massive stars and/or their remnants in subgalactic halos
that generated the ultraviolet radiation and mechanical energy that
reheated and reionized most of the hydrogen in the cosmos. The epoch
of reionization sets the stage for the subsequent evolution of stars,
galaxies and the intergalactic medium (IGM) and is currently the focus
of considerable observational and theoretical efforts.

The Lyman~$\alpha$ (\lya) forest properties appear to change rapidly
around $z \simeq 6$ (Fan et al. \citeyear{Fan06}): the mean length of
complete dark gaps (where the QSO light is completely absorbed)
increases; the variance in the forest properties increases, indicating
larger spatial fluctuations in the ultraviolet ionizing background;
and the sizes of the QSOs own H\,{\sc ii} regions decrease. Fan et
al. concluded that $z \sim 6$ marks the end of the overlapping stage
of reionization and that the mass-averaged neutral fraction of the IGM
is between 1 and 4\% at $z = 6.2$.
However, at present the comoving space density of
massive, luminous galaxies and QSOs at these epochs appears
insufficient to provide the required ionizing flux (e.g. Madau,
Haardt, \& Rees 1999; Fan et al. 2001; Bunker et al. 2004; Yan et
al. 2006). The recent third-year Wilkinson Microwave Anisotropy Probe
(WMAP) measurement of the optical depth to Thomson scattering,
$\tau_e=0.09\pm0.03$ (Spergel et al. \citeyear{Spergel06}) suggests
that the Universe was reionized at higher redshifts, an indication of
significant star formation activity at very early times.

The epoch of reionization is clearly a pivotal time in the history of
the Universe. Despite the recent progress described above, data are
still rather scarce. In particular, the properties of the \lya\ forest
can only be determined in a statistical way at these high redshifts,
since the line density is so high that individual \lya\ lines can no
longer be resolved.  In this regime, the metal lines which fall
longwards of the \lya\ emission take on a special significance as the
only tool at our disposal to recognise individual absorption systems,
whether in galaxies or the IGM, and probe the metal enrichment of the
Universe following the earliest episodes of star formation.

The \civ\ doublet ($\lambda\lambda$ 1548.2041, 1550.7812) is the most
common pair of metal lines associated with the highly ionized \lya\
forest (Sargent, Boksenberg, \& Steidel 1988). Metals may be
transported to the IGM in large-scale outflows (`superwinds') from
nearby actively star-forming galaxies, as proposed by Adelberger et
al. (2005). Alternatively, the IGM could have been enriched by an
early population of stars prior to the epoch of reionization
(e.g. Madau, Ferrara, \& Rees 2001), preferentially in regions around
the highest matter overdensities (Porciani \& Madau 2005).  The first
study to measure the amount of \civ\ in the IGM over a large range in
redshift was conducted by Songaila (2001) using a sample of 32 QSOs
observed with the Keck telescopes; a further analysis was presented in
Songaila (2005).  The somewhat surprising result of this work was the
finding that the cosmological mass density of C$^{3+}$ ions,
$\Omega_{\rm C\,IV}$, shows little, if any, redshift evolution between
$z = 1.5$ and 4.5.  Initially there was tentative evidence for a
turn-down of $\Omega_{\rm C\,IV}$ at $z \gtrsim 4.5$.  However, with
improved statistics Pettini et al. (2003) later showed the turn-over
to be probably due to incompleteness effects -- to the best of our
knowledge $\Omega_{\rm C\,IV}$ remains approximately constant as we
look back in time up to redshift $z \sim 5$.

Searches for \civ\ at redshifts beyond $z = 5$ are essential for
constraining the origin of metals in the IGM, since this is the
redshift where the comoving star formation rate density appears to
begin to decline (Bunker et al. 2004; Giavalisco et al. 2004; Bouwens
\& Illingworth 2006).  If the mass density of metals in the IGM drops
by a similar factor, then a scenario where winds from massive
star-forming galaxies pollute the IGM with metal will be favoured. On
the other hand, if the mass density of metals were observed to remain
constant, this may point to an epoch of very early enrichment of the
IGM, presumably by smaller mass objects.

The $z \sim 5$ limit of previous studies was set by the use of optical
spectrographs and CCD detectors -- the observations by Songaila (2001)
and Pettini et al. (2003) were all obtained with optical instruments,
the High Resolution Echelle Spectrograph and the Echelle Spectrograph
and Imager on the Keck telescopes.  At $z > 5$ the \civ\ doublet moves
to wavelengths beyond 9300\,\AA, requiring spectroscopy in the
near-infrared (IR).  This is a largely unexplored wavelength regime in
QSO absorption line spectroscopy (with a few exceptions -- see for
example Kobayashi et al. 2002).  On the other hand, near $1 \,\mu$m
there are several windows relatively free of terrestrial water vapour
absorption where, with suitably long exposure times, it should be
possible to achieve the signal-to-noise ratios and spectral resolution
required for narrow absorption line work.

To this end, we have conducted a pilot investigation with the Infrared
Spectrometer And Array Camera (ISAAC) on the ESO VLT (Moorwood et
al. 1998).  We targeted two of the highest redshift QSOs discovered by
the Sloan Digital Sky Survey: SDSS J103027.10+052455.0 ($z_{\rm
em}=6.28$, $J=18.87$, abbreviated to J1030+0524) and SDSS
J130608.26+035626.3 ($z_{\rm em}=5.99$, $J=18.77$, abbreviated to
J1306+0356) over the wavelength range 9880 to 10800\,\AA.  Details of
the observations and data reduction are given in Section~2.  We report
the detection of a definite, strong, \civ\ doublet at $z_{\rm abs} =
5.7238$ and of a possible doublet at $z_{\rm abs} = 5.8290$.  In
Section~3 we describe the measurement of the column densities of these
absorbers, $N_{\rm C\,IV}$, together with the derivation of the mass
density of C$^{3+}$ ions, $\Omega_{\rm C\,IV}$, they imply.  We
discuss these results in Section~4 and consider future prospects for
this work in Section~5.


\section{Observations and data reduction}


\begin{figure*}
\vspace{-5.05cm}
\centerline{\hspace{0.5cm}
\includegraphics[width=2\columnwidth,clip,angle=270]{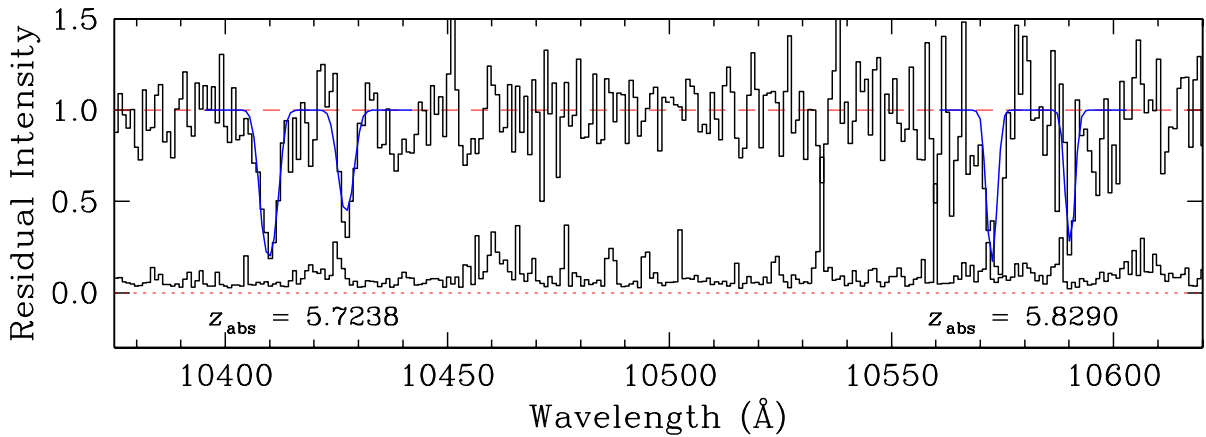}}
\vspace{-6cm}
\caption{Portion of the ISAAC spectrum of J1030+0524 containing the
C\,{\sc iv} absorption line system at $z_{\rm abs} = 5.7238$ and
possible \civ\ doublet at $z_{\rm abs} = 5.8290$. The top histogram
shows the normalised QSO spectrum, while the lower histogram is the
corresponding $1 \sigma$ error spectrum.  The continuous (blue) line
is the VPFIT model profile fitted to each C\,{\sc iv} doublet with the
parameters listed in Table~1.}
\medskip
\label{fig:spec1}
\end{figure*}

The ISAAC observations of J1030+0524 and J1306+0356 were obtained in
service mode in March 2004 (J1306+0356) and in the period
January--March 2005 (J1030+0524). ISAAC was used in its medium
resolution spectroscopic mode with a 0.6\,\arcsec\ slit to deliver a
resolving power $R \simeq 5700$ (FWHM\,$ \simeq 53$\,km~s$^{-1}$ or
1.8\,\AA\ at the wavelengths of interest here) sampled with four
0.146\,\arcsec\ ($\simeq 13$\,km~s$^{-1}$, 0.45\,\AA) pixels of the
Rockwell detector.  Our proposal was to use two wavelength settings,
each spanning a wavelength range $\Delta \lambda = 460$\,\AA, centred
respectively at $\lambda_{\rm c} = 10110$\,\AA\ and 10570\,\AA.  At
the conclusion of the service observations, a total of two hours of
data at the 10110\,\AA\ setting had been secured for J1306+0356. In
the case of J1030+0524, the total exposure time was 11 hours at
10110\,\AA\ and 10 hours at 10570\,\AA. For the latter object, two
additional hours were spent unintentionally at an intermediate setting
centred at 10330\,\AA; however, these data turned out to be useful too
as they provide an independent confirmation of the reality of the
strong \civ\ system at $z_{\rm abs} = 5.7238$ which is detected in
both wavelength settings even though it was recorded at different
physical locations on the detector.


\begin{table}
\begin{minipage}{105mm}
\caption{Parameters of identified \civ\ absorption systems}
  \begin{tabular}{llll} \hline QSO &
   $z_{\rm abs}$ & $\log (N_{\rm C\,IV}$/\cm) & $b$ (\kms) \\ 
  \hline 
  J1030+0524 & 5.7238 $\pm$ 0.0001 & 14.37 $\pm$0.05 & 63 $\pm$ 5\\ 
  J1030+0524 & 5.8290 $\pm$ 0.0002 & 14.5: (marginal) & ... \\ 
  \hline
\end{tabular}
\end{minipage}
\end{table}

The observations were performed in the conventional beam-switching
mode, with the QSO moved between two positions on the slit separated
by 10\,\arcsec.  After an A-B-B-A series of $4 \times 900$\,s long
exposures, the object was reacquired at a different position along the
slit (so as to use different pixel rows on the detector to record the
signal) and the four-exposure sequence repeated.  The 120\,\arcsec\
long ISAAC slit was oriented on the sky to include a bright star in
each observation, with the centre of the slit positioned approximately
half way between the bright star and the faint QSO. The bright star
aids target acquisition and provides a record of the seeing and sky
transparency during the observations; the weather conditions were
similar over most of our observations, with the seeing typically at
FWHM\,$\lesssim 1$\,\arcsec.

The data reduction used as a starting point the pipeline processed
two-dimensional images provided by ESO.  Each of these images
corresponds to a $4 \times 900$\,s A-B-B-A sequence where the
individual 900\,s exposures have been flat-fielded, combined so as to
subtract the background, and mapped to a geocentric, vacuum wavelength
scale. Our first step was to shift each 2D pre-processed image in the
spatial direction, using the bright star for reference, so that all
the exposures at a given wavelength setting are coaligned spatially;
only integer pixel shifts were applied.  Comparison of the OH sky line
residuals between different images revealed that the pipeline
wavelength calibration was only accurate to about $\pm 5$\,\AA; we
remedied this situation by applying a new dispersion correction using
the OH line list by Rousselot et al. (2000) and in the process changed
to a heliocentric wavelength scale.\footnote{Fortunately, the
heliocentric correction was not very different for different exposures
taken at the same wavelength setting, so that this remapping of the
wavelength scale did not lead to significant smearing of the sky line
residuals to be subtracted at a subsequent stage of the data
processing, as explained below.}

The properly registered 2D images were co-added using the IRAF task
{\sc imcombine}. In cases where more than two images are available at
the same wavelength setting, we used a `minmax' algorithm to reject
pixels affected by cosmic-ray events or other defects. When this was
not possible, such defective pixels were identified visually and
patched with the IRAF task {\sc fixpix} which interpolates between
adjacent areas on the image. The fully processed 2D images contain
three spectra of each object on the slit -- a positive spectrum
flanked by a negative spectrum on either side -- as a result of the
way the initial A-B-B-A exposures are combined. We extracted
one-dimensional spectra from the 2D images using the IRAF task {\sc
apall} on the central, positive, spectrum; in the extraction process
we also corrected for any residual sky emission by using standard
background subtraction tools available within {\sc apall}.

In cases where several exposures at the same wavelength setting are
available, we used the same {\sc apall} parameters to extract 1D
spectra from the individual 2D images (rather than from the co-added
one), and then co-added the 1D extractions to produce a final
spectrum. While the result of this alternative procedure is not
significantly different from the 2D addition, it does have the
advantage of providing an empirical measure of the associated 1D error
spectrum from consideration of the rms fluctuations in each wavelength
bin from its (mean) value in the final, co-added, spectrum.

In the last step, the 1D spectra were rebinned by a factor of two in
wavelength to improve the signal-to-noise ratio while maintaining the
sampling at the Nyquist limit. In the case of J1030+0524, we also
co-added the portion of the final spectra which is in common between
the two settings at central wavelengths $\lambda_{\rm c} =
10570$\,\AA\ and $\lambda_{\rm c} = 10330$\,\AA\ respectively.

\section{Intergalactic \civ\ at redshifts 5.4 -- 6}

The spectra were searched by eye for absorption features separated by
$\delta \lambda \times (1+z)$, where $\delta \lambda = 2.5771$\,\AA\
is the rest-frame wavelength difference between the two members of the
\civ\ doublet. An additional requirement for the identification of
such features as redshifted \civ\ lines is that their equivalent
widths should satisfy the condition $ 1 \leq W_{1548}/W_{1550} \leq
2$.  No such features could be recognised in the wavelength interval
$\Delta \lambda = 9880 - 10340$\,\AA\ (centred at $\lambda_{\rm c} =
10110$\,\AA) in the spectrum of either QSO.  However, two \civ\
doublets, one definite and one possible, were flagged in the second
spectrum of J1030+0524 which extends from 10340 to 10800\,\AA.  As
already explained, the first doublet, at $\lambda\lambda_{\rm obs} =
10409.8, 10427.1$\,\AA, is also present in the spectrum of J1030+0524
centred at $\lambda_{\rm c} = 10330$\,\AA, providing additional
confirmation that it is indeed real absorption rather than a detector
defect or an artifact of the data processing.

The two \civ\ doublets are shown in Figure~\ref{fig:spec1} after
normalisation by the underlying QSO continuum. In order to deduce the
parameters of the absorbers, we fitted the absorption lines with
theoretical Voigt profiles generated by the VPFIT software
package\footnote{VPFIT is available at
http://www.ast.cam.ac.uk/$\sim$rfc/vpfit.html\,.}; the values of
redshift $z_{\rm abs}$, \civ\ column density $\log N_{\rm C\,IV}$, and
Doppler parameter $b$ (and their errors) returned by VPFIT are
collected in Table~1. A single absorption component was fitted in each
case -- a more complicated velocity structure, while plausible, is
unwarranted by the quality of the present data.

The line fitting procedure shows that the $z_{\rm abs} = 5.8290$
system is only a tentative identification. As can be seen from
Figure~1, the two absorption features are not centred at exactly the
same redshift, although the wavelength mismatch could be due to the
noisy character of the spectrum.  Better data will be required to
confirm the reality of this absorber---for the moment, we flag it as
`possible'. Observations taken with the Gemini Near-IR Spectrograph
(GNIRS) on Gemini South (Simcoe 2006) also show absorption at $z_{\rm
abs}=5.8290$. In addition, two weak \civ\ systems identified in the
GNIRS data appear as marginal features in our ISAAC data at
$z_{\rm abs}=5.46$ and 5.74 respectively.

On the other hand, VPFIT confirms that the pair of absorption lines at
$\lambda\lambda_{\rm obs} = 10409.8, 10427.1$\,\AA\ is well fitted by
a \civ\ doublet at $z_{\rm abs} = 5.7238$ (see Figure~1). This is a
strong absorber, with $N_{\rm C\,IV}= 2.35 \times 10^{14}$\,cm$^{-2}$,
higher than any of the values of $N_{\rm C\,IV}$ found by Pettini et
al. (2003) and Songaila (2001) at $z_{\rm abs} = 4.5 - 5$. The
absorption lines are wide and fully resolved---the instrumental
resolution of our ISAAC spectra corresponds to a Doppler parameter
$b_{\rm INSTR} = 32$\,km~s$^{-1}$, much smaller than the value $b_{\rm
C\,IV} = 63 \pm 5$\,km~s$^{-1}$ returned by VPFIT for the $z_{\rm abs}
= 5.7238$ doublet (after deconvolution with the instrumental
broadening).  If such wide absorption profiles result from the
superposition of multiple velocity components of smaller internal
velocity dispersion, we may have underestimated the total column
density of \civ\ absorbers and the value returned by VPFIT would in
that case be a lower limit to $N_{\rm C\,IV}$.

Clearly, it would be unwise to draw sweeping conclusions on the basis
of a single detection. On the one hand, the strong $z_{\rm abs} =
5.7238$ \civ\ system we have discovered in the spectrum of the $z_{\rm
em} = 6.28$ QSO J1030+0524 may not be typical of the IGM at these
redshifts (although the tentative detection of a second strong
absorber, if confirmed, would argue against this interpretation).  On
the other, one could argue that the one detection reported here gives
a lower limit to the density of intergalactic \civ\ because: (i) the
absorption lines may be saturated and $N_{\rm C\,IV}$ may have been
underestimated, as explained above; (ii) a second \civ\ system of
comparable column density may have been detected; and (iii) our data
are only sensitive to the strongest \civ\ absorbers and miss the more
common lower column density systems. The marginal features 
discussed above suggest that deeper observations are indeed
likely to reveal more \civ\ absorbers that could make a substantial 
contribution to the total mass
density of \civ. 

Let us denote with $\Delta X$ the total absorption distance sampled by
our spectra; for ease of comparison with earlier work, we adopt an
Einstein-de Sitter cosmology in which the absorption distance is given
by:
\begin{equation}
\Delta X =  \frac{2}{3}
\sum_{j} [(1+z^{\rm max}_j)^{3/2}-(1+z^{\rm min}_j)^{3/2}]\, ,
\end{equation}
where the summation is over $j$ spectra in which \civ\ doublets could
have been detected at redshifts between $z^{\rm max}_j$ and $z^{\rm
min}_j$.  Although the S/N ratio of our spectrum of J1306+0356 is
lower than that of either of the two spectral regions recorded in
J1030+0524, it is nevertheless sufficient to detect \civ\ absorbers as
strong as those shown in Figure~\ref{fig:spec1}. Thus, in the present
analysis, $j = 3$ and $\Delta X = 2.005$. (The effective pathlength is
less than $3\times460$\,\AA, since we have taken into account the
minimum and maximum redshifts at which both components of the doublet
would be detected and have not included portions of the spectra with
significant skyline residuals or other lines).

Songaila's best fit to the $f(N)$ distribution in the redshift range
$2.90 < z <3.54$ is $f(N)=10^{-12.4} N_{13}^{-1.8}$, where $N_{13}$ is
the \civ\ column density measured in units of $10^{13}\,$cm$^{-2}$. If
this steep slope of $f(N)$ were to persist to the higher redshifts
probed here, by integrating over the column density we would expect
$f(>N)=5\,N_{13}^{-0.8}$, that is about 0.5 absorbers with $N \geq
2\times10^{14}$ \cm\ per unit absorption distance.  The detection of
one absorber with $N_{\rm C\,IV} \geq 2\times 10^{14}$ \cm\ over an
absorption distance of $\Delta X =2$ is therefore in agreement with
the expected value.  Within the limited statistics of our sample, it
appears that there is little evolution in the density of \civ\ systems
with $N_{\rm C\,IV} \geq 2\times10^{14}$ \cm.

\begin{figure}
\vspace{13pc}
\includegraphics{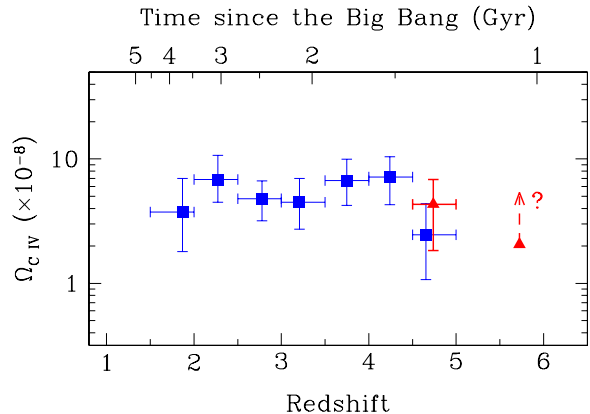}
\caption{The comoving density of C\,{\sc iv} in units of the critical
density, calculated in a $(\Omega_{\rm
M},\Omega_\Lambda,h)=(1,0,0.65)$ cosmology for ease of comparison with
earlier work. (Note, however, that the timescale given on the top axis
is appropriate to today's consensus cosmology with $\Omega_{\rm
M},\Omega_\Lambda,h = 0.3,0.7,0.65$.)  The blue squares show the
measurements by Songaila (2001), while the red triangles are the high
redshift values determined by Pettini et al. (2003) and here. The
difference in $\Omega_{\rm C\,IV}$ in the interval $z_{\rm abs} = 4.5
- 5.0$ between Songaila's and Pettini et al.'s estimates is due
primarily to the incompleteness correction applied to the latter
dataset. The newly determined value at $z_{\rm abs}\sim 5.7$ is likely
to be a lower limit for the reasons explained in the text.}
\label{fig:omega}
\end{figure}

We can calculate the mass density of C$^{3+}$ ions implied by our
detection using the usual statistic (Pettini et al. 2003):
\begin{equation}
\Omega_{\rm C\,IV}=\frac{H_0 m_{\rm C\,IV}}{c\rho_{\rm crit}} \frac{\sum
N_{\rm C\,IV}}{\Delta X} \, ,
\label{eqn:omega1}
\end{equation}
\noindent where $m_{\rm C\,IV}$ is the mass of a \civ\ ion,
and  $\rho_{\rm crit}$ is
the critical density. Entering the numerical values
and with $H_0 = 65$\,km~s$^{-1}$~Mpc$^{-1}$, 
eq.\,(\ref{eqn:omega1}) reduces to:
\begin{equation}
\Omega_{\rm C\,IV}=1.75\times10^{-22}\frac{\sum N_{\rm C\,IV}}{\Delta X}, 
\label{eqn:omega2}
\end{equation}
leading to $\Omega_{\rm C\,IV} = 2.1 \times10^{-8}$ at $z \simeq 5.7$.
We have plotted this value as a lower limit (for the reasons given
above) in Figure~\ref{fig:omega}. Comparison with analogous
measurements at lower redshifts shows that, \emph{if typical of the
widespread intergalactic medium}, the value of $\Omega_{\rm C\,IV}$
deduced here is consistent with little, if any, decrease in this
quantity from $z = 1.5$ to the highest redshifts that can be probed
with known QSOs. Specifically, the lower limit we have determined in
the present work is very similar to Songaila's (2001) first estimate
of $\Omega_{\rm C\,IV}$ at $z = 4.5 - 5$ which also based on limited
statistics.  Deeper observations by Pettini et al. (2003) later showed
that incompleteness corrections raise this value to approximately the
same level as measured at $z < 4.5$, and this may well be the case too
for the data reported here.


\section{Discussion}

The apparent lack of evolution of the quantity $\Omega_{\rm C\,IV}$
from $ \sim 1$ to $\sim 4$\,Gyr after the Big Bang is puzzling indeed,
given the dramatic changes we see in many other properties of the
Universe at these epochs.  Taken at face value, the approximately
constant value of $\Omega_{\rm C\,IV}$ from $z = 1.5$ to $\sim 6$
would seem to argue for an early ($z > 6$) production of intergalactic
metals, consistent with the interpretation put forward by Madau et
al. (2001) and Porciani \& Madau (2005).  An alternative explanation,
favoured by Adelberger et al. (2005; see also Adelberger 2005), is
that a significant fraction of the `intergalactic' \civ\ absorption
actually occurs in the vicinity of actively star-forming galaxies
undergoing mass loss on a galaxy-wide scale as a direct consequence of
the star formation activity we see.  The statistical association of
strong \civ\ absorption systems with galaxies close to the QSO
sightline at $z = 2 - 3$ in the surveys by Steidel and collaborators
(Steidel et al. 2004) and the ubiquitous evidence for mass outflows
from these galaxies (Pettini et al. 2001; Steidel et al. 2004; Erb et
al. 2006) support this picture.

In the latter case, the total amount of carbon in the IGM,
$\Omega_{\rm C}$, would presumably increase with the progress of time
as a result of the on-going star formation activity in galaxies.
Given that only a fraction of carbon is triply ionized, even under the
physical conditions most favourable to C$^{3+}$, we would then have to
conclude that changes in the ion fraction C$^{3+}$/C$_{\rm TOT}$ and
$\Omega_{\rm C}$ work in opposite directions to produce the
approximately constant value of $\Omega_{\rm C\,IV}$ revealed by the
observations.  Oppenheimer \& Dav\'e (2006) have recently investigated
such a scenario with hydrodynamic simulations of structure formation
which incorporate outflows from star forming galaxies and shown it to
be plausible.

The lack of C\,{\sc ii}~$\lambda 1334$ absorption at $z=5.7238$ in the
optical spectrum of J1030+0524 (Pettini et al. 2003) indicates that
the \civ\ we detect arises in highly ionized, optically thin gas, as
is the case for most of the \civ\ absorbers at lower redshifts.
Conversely, Becker et al. (2006) recently reported the detection of a
surprisingly large number of O\,{\sc i} absorption systems at $z_{\rm
abs} > 5$ in the spectrum of the $z_{\rm em} = 6.42$ QSO SDSS
J1148+5251 and speculated that a decrease in high ionization species
such as \civ, and corresponding rise in low ionization species such as
O\,{\sc i}, may signal a shift in the ionization state of the IGM at
$z>5$.  The results reported here suggest that the real situation is
probably more complicated, in that the evidence for a widespread
decrease in $\Omega_{\rm C\,IV}$ at $ z > 5$ is far from secure at
present, as we have argued.  One difficulty of course is `cosmic
variance' -- without larger samples we do not know that the sightline
to J1030+0524 does not exhibit an excess of strong \civ\ systems, just
as the sightline to J1148+5251 shows an unusually high number of
O\,{\sc i} absorbers.

Is there any evidence that the strong \civ\ absorber at $z_{\rm
abs} = 5.7238$ in J1030+0524 is associated with outflowing gas
from a foreground galaxy?  Songaila (2006) has proposed that about
half of all high column density ($N_{\rm C\,IV}>2\times10^{13}$ \cm)
\civ\ systems, which have a median full width at one-tenth maximum
velocity of FWTM$_{10\%} = 160$\,\kms\ could originate in galactic 
outflows. The $z_{\rm abs} = 5.7238$ absorber in J1030+0524
certainly fits into this category since the value of $b =
63$\,km~s$^{-1}$ we measure corresponds to FWTM$_{10\%} = 191$\,\kms.
Galaxies at $z \sim 6$ are known to contain metal enriched
interstellar gas which is seen in absorption against the central
starburst (e.g. Taniguchi et al. 2005).  Low ionization metal
absorption lines are also detected in the host galaxy of a
$\gamma$-ray burst (GRB) at $z=6.3$, possibly the result of a
metal-enriched nebula swept up by a progenitor wind prior to the GRB
(Kawai et al. \citeyear{Kawai06}).  It is also likely that massive
galaxies at $z \sim 6$ support significant outflows, since their
rest-frame UV luminosities indicate star formation rates in the range
3 to 30\,\msun~yr$^{-1}$ (Yan et al. \citeyear{Yan06}).

Interestingly, the field of the QSO J1030+0524 has been imaged with
the Advanced Camera for Surveys on the \emph{Hubble Space Telescope}
by Stiavelli et al. (2005). These authors found an excess of objects
with the colours of galaxies at $z \gtrsim 5.5$ near the sightline to
the QSO; two are located 85 and 89 proper kpc ($\Omega_{\rm
M},\Omega_\Lambda,h = 0.3,0.7,0.65$) from the QSO sightline,
respectively, but their redshifts are unknown. For comparison,
Adelberger et al. (2005) found that about half of the $z \simeq 3$
Lyman break galaxies in their sample produce strong ($N_{\rm
C\,IV}\sim 10^{14}$ \cm) \civ\ absorption within $\sim 80$ kpc of the
galaxy. The only galaxy in the Stiavelli et al. (2005) sample with a
spectroscopically confirmed redshift ($z=5.970$)---determined from its
Ly$\alpha$ emission line which, incidentally, exhibits the asymmetric
P-Cygni profile characteristic of outflowing gas---is located 350
proper kpc from the QSO sightline, beyond the projected distance where
Adelberger et al. (2005) typically find a significant excess of \civ\
absorbers. The four remaining $z \gtrsim 5.5$ galaxies are also
located beyond this projected distance. It may be the case that the
excess of galaxy counts is indeed evidence for a large-scale
overdensity (as proposed by Stiavelli et al. 2005), where star
formation and metal enrichment started at earlier times (Steidel et
al. 2005).  The fact that four of the six \oi\ systems found by Becker
et al. (2006) are along one line of sight may also suggest a tendency
for patchy, and thus highly biased, metal enrichment at high redshift.

\section{Summary and Outlook}

In summary, we have detected one definite and one marginal \civ\
absorption line system between $z_{\rm abs} = 5.40$ and 6.0\,.  These
are the highest redshifts at which \civ\ has been detected to
date. The column density we measure in the one definite system at
$z_{\rm abs} = 5.7238$ gives a lower limit to the comoving mass
density of C$^{3+}$ ions at these redshifts $\Omega_{\rm C\,IV} \geq
2.1 \times 10^{-8}$; this value is consistent with only mild or no
evolution of $\Omega_{\rm C\,IV}$ from $z<5$.  The finding that a
substantial fraction of intergalactic metals is already in the IGM at
the highest redshifts that can be probed with known QSOs provides
tantalising evidence for an early epoch of metal production.  However,
with the detection of just one absorption system, such a conclusion
would be premature.  Finding patchy enrichment in matter overdensities
can be used to argue either case: early enrichment by stars at the
onset of reionization or later injection of metals into the IGM from
starburst winds.

Extending the search for metals to lower column density systems,
coupled with deep imaging and spectroscopy of galaxies near high-$z$
QSO sightlines, would help discriminate between these two scenarios.
Having demonstrated that QSO absorption line spectroscopy can be
successfully performed at near-IR wavelengths, we now plan to increase
the pathlength of our survey with observations towards other $z_{\rm
em} \sim 6$ QSOs.  Looking further ahead, the forthcoming availability
of X-shooter on the VLT (Moorwood \& D'Dodorico 2004), a second
generation instrument which is much better suited to the requirements
of QSO absorption line spectroscopy than ISAAC, promises to throw wide
open the observational window on stellar nucleosynthesis during the
first Gyr of the Universe.

We thank Xiaohui Fan for providing coordinates of the two QSOs prior
to their publication and Massimo Stiavelli for communicating the
positions of galaxies in the J1030+0524 field. We are grateful to the
VLT time assignment committee for the award of ISAAC time and to the
ESO support astronomers who carried out the observations on our
behalf.  Financial support for this work was provided by a PPARC
rolling grant at the University of Cambridge (E.R-W.), and by NSF
grant AST02-05738 and NASA grant NNG04GK85G (P.M.).

\bibliographystyle{mn2e}
\bibliography{mn-jour,civ}

\bsp

\label{lastpage}

\end{document}